\documentstyle{basi}
\begin{document}                                                               
\title[Helicity in dynamos]{Magnetic helicity in galactic dynamos} 
\author[Kandaswamy Subramanian]
{Kandaswamy Subramanian\thanks{e-mail:kandu@iucaa.ernet.in} \\                 
        Inter University Centre for Astronomy and Astrophysics, \\             
        Post bag 4, Pune University campus, Ganeshkhind, Pune 411 007}         
\maketitle
\label{firstpage}
\begin{abstract}
Magnetic fields correlated on kiloparsec scales are seen
in spiral galaxies. Their origin could be due to amplification
of a small seed field by a turbulent galactic dynamo.
We review the current status of the galactic dynamo,
especially the constraints imposed by magnetic helicity conservation.
We estimate the minimal strength of the large-scale magnetic field
which could arise inspite of the helicity constraint.
\end{abstract}

\section{The galactic dynamo}

Magnetic fields in spiral galaxies have strengths of order few $10^{-6} G$,
and are coherent on scales of several kpc (Beck et al. 1996).
In several disk galaxies, like M51 and NGC 6946, they are
also highly correlated (or anti-correlated)
with the optical spiral arms.
How do such ordered, large-scale fields arise?
One possibility is the dynamo amplification of a weak but nonzero seed
field. We critically review here the operation of the
galactic dynamo, particularly emphasing the constraints which arise due
to the conservation of magnetic helicity in highly conducting plasma.

The evolution of the magnetic field is described by the induction equation
\begin{equation}
{\partial {\bf B} \over \partial t} =
{\bf \nabla } \times ( {\bf v} \times {\bf B} -
 \eta {\bf \nabla } \times {\bf B}).
\label{induc}
\end{equation}
Here ${\bf B}$ is the magnetic field,
${\bf v}$ the velocity of the fluid and
$\eta$ the resistivity.
${\bf B} =0$ is a perfectly valid solution of the induction
equation. So there would be no magnetic field generated if
one were to start with a zero magnetic field.
There are a number of battery mechanisms, 
invoking small additional source terms 
to Ohms law, which lead to a seed magnetic field from
zero fields (cf. Rees 1994;
Subramanian, Narasimha and Chitre 1994).
This seed field is generically
much smaller than the galactic fields. Therefore some form of
dynamo action, due to motions which act to exponentiate
small seed fields efficiently, is essential to explain observed
galactic fields.

Galactic dynamos depend on the following two features: First,
disk galaxies are differentially rotating systems.
Also the magnetic flux is to a
large extent frozen into the fluid. So any radial
component of the magnetic field will be efficiently wound
up and amplified to produce a toroidal component.
But this results in only a linear amplification of the field.
To obtain the observed
galactic fields starting from small seed fields
one needs a way to generate the radial
component from the toroidal one.
If this can be done, the field can grow exponentially and
one has a dynamo.

A mechanism to produce the radial field from the toroidal
field was invented by Parker (1955), and is known
as the $\alpha$-effect (Steenbeck, Krause and Radler 1966).
The essential feature is to invoke the effects of cyclonic 
turbulence in the galactic gas (cf. Ferriere 1998). 
The interstellar medium is assumed
to be turbulent, due to for example the effect of supernovae randomly
going off in different regions. In a rotating, stratified
(in density and pressure) medium
like a disk galaxy, such turbulence becomes helical.
An upward moving fluid parcel, expands and the coriolis
force makes it rotate retrogade, generating negative
kinetic helicity in the northern hemisphere.
Downward moving fluid contracts, and the coriolis force now
makes it rotate in the prograde direction. This contributes
to helicity of the same sense.
Helical motions of the gas
perpendicular to the disk draws out the toroidal field
into a loop which looks like a {\it twisted} $\Omega$.
Such a twisted loop is connected to a current
which has a component parallel to the original toroidal 
field. If the motions
have a non-zero net helicity, this
parallel component of the current adds up coherently.
A toroidal current then results from the toroidal field.
Hence, poloidal fields can be generated from toroidal ones.
(Of course microscopic diffusion is essential
to make permanent changes in the field).
This closes the toroidal-poloidal cycle and leads
to exponential growth of the mean field.

In quantitative terms, suppose the velocity field
is the sum of a mean, large-scale velocity ${\bf V}_0$
and a turbulent, stochastic velocity ${\bf v}_T$.
The induction equation becomes a stochastic
partial differential equation. 
Split the magnetic field ${\bf B} = {\bf B}_0
+ {\bf b}$, into a mean field $ {\bf B}_0 = < {\bf B}>$ and a
fluctuating component ${\bf b}$. Here the average $<>$, is defined
either as a spatial average over scales larger than the turbulent
eddy scales (but smaller than the system size) or as an ensemble average.
Assume the turbulence to be isotropic, homogeneous, helical
and have a short (ideally delta function) correlation
time $\tau$. Then one can
derive the mean-field dynamo equation for ${\bf B}_0$,
\begin{equation}
{\partial {\bf B}_0 \over \partial t} =
{\bf \nabla } \times \left( {\bf V}_0 \times {\bf B}_0 +
{\bf \varepsilon} - \eta {\bf \nabla } \times {\bf B}_0 \right) .
\end{equation}
\begin{equation}
{\bf \varepsilon} = < {\bf v}_T \times {\bf b} >
= \alpha_0 {\bf B}_0 - \eta_T {\bf \nabla } \times {\bf B}_0.
\end{equation}
Here ${\bf \varepsilon}$ is the turbulent emf,
$\alpha_0 = -(\tau/3) < {\bf v}_T.({\bf \nabla} \times {\bf v}_T) > $,
is the dynamo $\alpha$-effect, proportional to the
kinetic helicity and $\eta_T = (\tau/3) < {\bf v}_T^2 > $,
is the turbulent magnetic diffusivity proportional to the kinetic energy
of the turbulence. This kinematic mean-field dynamo equation,
has exponentially growing solutions, provided
a dimensionless dynamo number has magnitude
$D = \vert \alpha_0 G h^3 \eta_T^{-2} \vert > 
D_{crit} \sim 6$ (Ruzmaikin, Shukurov and Sokoloff 1988).
(Here $h$ is the disk scale height and $G$ the galactic shear,
and we have defined $D$ to be positive).
While the $\alpha$-effect is crucial for regeneration
of poloidal from toroidal fields, the turbulent diffusion
turns out to be also essential for allowing changes
in the mean field flux. The mean field grows
typically on time-scales a few times the rotation time scales,
of order $10^9$ yr. Modulations of $\alpha$, and $\eta_T$
due to the spiral arms, can also lead to
large-scale fields, correlated (or anti-correlated)
with the optical spirals (Mestel and Subramanian 1991; 
Moss 1998; Shukurov 1998).

The kinematic mean-field equation neglects the
the back-reaction on the velocity due to the Lorentz forces.
This rapidly becomes a bad approximation, due
to the more rapid build up of magnetic noise
compared to the mean field (Kulsrud and Anderson 1992).
Both direct numerical simulations of the 
non-linear dynamo (Brandenburg 2001, Brandenburg and Sarson 2002,
Brandenburg, Dobler and Subramanian 2002) and semi-analytic modelling
of the non-linear effects (Subramanian 1999; Brandenburg and
Subramanian 2000) point to the crucial 
role played by magnetic helicity conservation in limiting 
mean field growth.

\section{ Magnetic helicity conservation and the galactic dynamo}

The magnetic helcity associated with a field
${\bf B} = {\bf \nabla } \times {\bf A}$ is defined as
$H = \int {\bf A}.{\bf B} \ dV$, where ${\bf A}$ is the vector
potential (Moffat 1978, Berger and Field 1984). Note that this definition of
helicity is only gauge invaraiant (and hence meaningful) if the domain
of integration is periodic, infinite or has
a boundary where the normal component of the field vanishes.
In this case, under a gauge transformation
${\bf A} \to {\bf A} - {\bf \nabla}\psi$, the additional term
in the helicity, $\int {\bf \nabla}\psi . {\bf B} =
\int \psi {\bf B}.d{\bf S} - \int \psi {\bf \nabla}.{\bf B} \ dV
= 0$. $H$ measures the linkages and twists in the magnetic field.
From the induction equation one can easily derive the
helicity conservation equation,
\begin{equation}
{dH \over dt} = -2 \eta \int {4\pi \over c} {\bf J}.{\bf B} \ dV,
\label{helcon}
\end{equation}
where ${\bf J} = (c/4\pi) {\bf \nabla} \times {\bf B}$ is the
current density. So in ideal MHD with $\eta = 0$, magnetic helicity
is strictly conserved. However, this does not guarantee
conservation of $H$ in the limit $\eta\rightarrow0$,
because the current helicity, $\int {\bf J}.{\bf B} \ dV$, may still
become large. For example, the Ohmic dissipation rate of magnetic energy
$Q_{\rm Joule}\equiv\eta (4\pi/c^2) \int {\bf J}^2 dV$
can be finite and balance magnetic energy input by motions,
even when $\eta \rightarrow0$. This is
because small enough scales develop in the field (current sheets)
where the current density increases with decreasing $\eta$ as
$\propto\eta^{-1/2}$ as $\eta\rightarrow0$,
whilst the rms magnetic field strength, $B_{\rm rms}$, remains
essentially independent of $\eta$. Even in this case, however,
the rate of magnetic helicity dissipation {\it decreases} with $\eta$
like $\propto\eta^{+1/2}\rightarrow0$, as $\eta\rightarrow0$.
Thus, under many astrophysical conditions where $R_m$ is
large ($\eta$ small), the magnetic helicity $H$, is almost independent
of time, even when the magnetic energy is dissipated at finite rates.

Coming back to the mean-field dynamo, we note that its
operation automatically leads to the growth of linkages between the
toroidal and poloidal mean fields. Such linkages measure the helicity
associated with the mean field. One then wonders how
this mean field (galactic) helicity arises?
To understand this, we need to split the helicity conservation
equation into evolution equations of the
sub-helicities associated with the mean field, say
$H_0 = \int {\bf A}_0.{\bf B}_0 \ dV$ and the fluctuating
field $h = \int <{\bf a}.{\bf b}> dV = <{\bf a}.{\bf b}> V$.
The evolution equations for $H_0$ and $h$ are
\begin{equation}
{dH_0 \over dt} =\int 2{\bf \varepsilon}.{\bf B}_0 \ dV
 -2 \eta \int {4\pi \over c} {\bf J_0}.{\bf B_0} \ dV
\label{H0con}
\end{equation}
\begin{equation}
{dh \over dt} =-\int 2{\bf \varepsilon}.{\bf B}_0 \ dV
 -2 \eta \int {4\pi \over c} <{\bf j}.{\bf b} > dV .
\label{hcon}
\end{equation}
Here, and henceforth, we assume that the surface terms can
either be neglected or are zero (becuase of boundry conditions).
We see that the turbulent emf ${\bf \varepsilon}$
transfers helicity between large and small scales;
it puts equal and opposite amounts of helicity into
the mean field and the small-scale field, conserving
the total helicity $H = H_0 + h$. So if one were to start with
zero total helicity intially, in a system with
large $R_m$, one will always have $H_0 + h \approx 0$,
or $\vert H_0 \vert \approx \vert h \vert$.

Note that for a given amount of helicity, the energy
associated with the field is inversely proportional
to the scale over which the field varies.
If for example, the small-scale field were 
maximally helical, and varied on a single scale, with associated
wave number, $k_f$, we will have $ k_f <{\bf a}.{\bf b}> = < {\bf b}^2 >$.
Similarly in a periodic box, a maximally helical
large scale field with wave number $k_m$, satisfies,
$k_m \int dV {\bf A}_0.{\bf B}_0 =\int dV {\bf B}_0^2 $.
(Henceforth, we will denote the volume average of
mean field quantities $X_0$ over the scale
of the system, $\int (dV/V) X_0$, by $\overline{X_0}$ ).
So, helicity conservation, with $\vert H_0 \vert \approx
\vert h \vert$, implies $\overline{{\bf B}_0^2}
\approx (k_m/k_f) < {\bf b}^2 >$.
Now in general, $< {\bf b}^2 >^{1/2}$, will saturate 
near the equipartition field strength, say
$B_{eq}^2 = 4\pi \rho {\bf v}_T^2$.
(Here $\rho$ is the fluid density).
So one obtains $ \overline{{\bf B}_0^2}
\approx (k_m/k_f) B_{eq}^2 \ll B_{eq}^2$, for
$k_m/k_f \ll 1$. The mean field is 
expected to attain atmost sub-equipartition values 
for $R_M\gg1$, if helicity is strictly conserved.

The galactic dynamo also involves shear and
the generation of the toroidal field by shear,
does not involve the generation of net helicity.
A periodic box simulation with an imposed periodic shear
(Brandenburg, Bigazzi and Subramanian 2001), suggests
that the helicity constraint applies now to the product of
the mean toroidal ($B_{t}$) and poloidal fields
($B_{p}$). So $ B_{t} B_{p}/k_m \approx 
\vert <{\bf a}.{\bf b}> \vert 
\approx \delta <{\bf b}^2>/k_f$. Here $\delta < 1$ takes 
into account that the small-scale field will also not be 
fully helical. In galaxies one usually has
$B_{t}/B_{p} = Q > 1$. This implies 
$B_{t}^2 \sim (Q\delta) (k_m/k_f) B_{eq}^2$.
In principal one can have large $B_{t}$ at the cost
of $B_{p}$, with strong shear.
These estimates give upper limits to the
mean-field strength with and without shear.
However, in both cases, the limits are smaller than 
$B_{eq}$ only by the {\it square root} of
the ratio of small to large scales (and by a further factor
$(Q\delta)^{1/2}$, in case of shear).
Whether such mean field strengths are indeed realised,
depends also on detailed dynamics of mean-field dynamo 
saturation, to which we now turn.

\section{ Modelling dynamo saturation and minimal mean galactic fields}

As a crude model of how the dynamo saturates, when the dynamo
is not too supercritical (see below), one may use
the quasi-linear theory applicable to weak mean fields
(cf. Pouquet, Frisch and Leorat 1976;
Zeldovich, Ruzmaikin and Sokoloff 1983; Gruzinov and Diamond 1994;
Bhattacharjee and Yuan 1995; Subramanian 2002a). 
This gives a re-normalised turbulent emf, with
$\alpha = \alpha_0 + \alpha_M$, where
$\alpha_M = (\tau/3) <{\bf b}.{\bf \nabla} \times {\bf b}>/(4\pi\rho)$,
is proportional to the small-scale current helicity. The turbulent difusion
$\eta_T$ is left unchanged to the lowest order, although to the next
order there arises a non-linear hyperdiffusive correction to
${\bf \varepsilon}$ (Subramanian 2002a).
One can now look for a combined steady state solution
to the helicity conservation equation (\ref{hcon}), and the mean-field
dynamo equation. This work is in progress (Subramanian 2000b),
and preliminary results are reported here.
(Similar work is also being done by Brandenburg and Blackman
(private communication); see also Brandenburg 2002;
 Field and Blackman 2002). Assume again that
the small-scale field has a scale $k_f^{-1}$. 
We then have $<{\bf b}.{\bf \nabla} \times {\bf b}> = 
k_f^2 < {\bf a}.{\bf b}>$ (although, for fields which
are not maximally helical, these helicities can not be related to the
energy). Using this, we can write $dh/dt$ in Eq. (\ref{hcon}),
in terms of $d\alpha_M/dt$ and hence in terms of $d\alpha/dt$.
Further, for $R_m$ large enough that helicity is conserved,
and for a large-scale field which varies on scale $k_m^{-1}$,
we can approximately write, 
$\overline{({\bf \nabla} \times {\bf B}_0).{\bf B}_0} \approx
k_m^2 \overline{ {\bf A}_0.{\bf B}_0}  = - k_m^2 < {\bf a}.{\bf b}>
= -(k_m/k_f)^2 <{\bf b}.{\bf \nabla} \times {\bf b}> $.
This relation is of course strictly valid only if
$H_0$ is well defined. This inturn requires
that negligible mean magnetic flux leave the disk,
which is likely in thin disk dynamos 
(cf. Ruzmaikin, Shukurov and Sokoloff 1988).
(A more careful treatment will involve using the
relative helicity of Berger and Field 1984).
The helicity conservation equation then gives a dynamical
equation for $\alpha$-quenching
\begin{equation}
{1\over \eta_T k_f^2} {d\alpha \over dt} =- 2 \alpha {\overline{{\bf B}_0^2}
\over B_{eq}^2} - 2{k_m^2 \over k_f^2} (\alpha - \alpha_0)
-2 {\eta \over \eta_T} (\alpha - \alpha_0)
\label{alpevol}
\end{equation}
We see that non-linear effects leads to a decrease in $\alpha$ with
time, till the RHS side Eq. (\ref{alpevol}) becomes zero.
(Such quenching of $\alpha$ has earlier been discussed by Kleeorin
and Ruzmaikin 1982; Zeldovich, Ruzmaikin and Sokoloff 1983).
As $\alpha$ decreases, the effective dynamo number of the
galactic dynamo, $D = \vert \alpha G h^3 \eta_T^{-2}\vert$, will also
decrease from an initial value $D_0 = \vert\alpha_0 G h^3 \eta_T^{-2}\vert$
and lead to a saturation of the mean field
growth when $D= D_{crit}$.
This happens when $\alpha = \alpha_{sat} = \alpha_0 (D_{crit}/D_0)$.
The stationary solution for {\it both} the dynamical $\alpha$ quenching
equation and the mean field dynamo equations is then
obtained by equating the RHS of Eq. (\ref{alpevol}) to zero,
and substituting a value of $\alpha = \alpha_{sat}$ given above.
For $\eta/\eta_T \ll (k_m/k_f)^2$, this gives
an estimated mean field strength $B_{mean} = \vert {\bf B}_0 \vert$
\begin{equation}
B_{mean} \approx [ (D_0/D_{crit}) - 1 ]^{1/2}
{k_m \over k_f} B_{eq}.
\label{minB}
\end{equation}
For galaxies $D_0 \sim 10 - 20$.
If we adopt $D_0/D_{crit} \sim 2$, $k_m/k_f \sim l/h$, where
$l \sim 100 pc$ is the forcing scale of the turbulence,
$h \sim 400 - 1000pc$,
then the mean field strenth would be $1/4$ to $1/10$ of
equipartition at saturation. This estimate
is more pessimistic than the more general limit 
$B_t < (Q\delta)^{1/2}(k_m/k_f)^{1/2} B_{eq}$, in section 2.
This is basically because, for $D_0$ exceeding but near $D_{crit}$,
the large-scale dynamo saturates even with  modest
$\alpha$ suppression, after
which there is no further helicity transfer ($\delta << 1$).
On the other hand, 
for $D_0 >> D_{crit}$ Eq. (\ref{minB})
seems to suggest fields greater than obtained
in section 2. This rather points to the limitations 
of the quasi-linear model for non-linear saturartion,
in this case, than to a violation of the more general limit.

A major caveat to the above limits is that 
galaxies have boundaries, and if
one has a flux of helicity due
to small-scale fields preferentially leaving the system
then one may avoid the above constraints 
(cf. Blackman and Field 2000; Kleeorin et al. 2000).
Artificial removal of small-scale fields
in a simulation, periodically, does indeed lead to
enhanced large-scale field growth; so the idea
works in principle (Brandenburg, Dobler and Subramanian 2002).
But so far the simulations which involve boundaries 
do not show a preferential out-flux of small-scale
field helicity (Brandenburg and Dobler 2001).
Another possibility is to think of a non-helical
dynamo (Vishniac and Cho 2001), but there
is no evidence yet for its working in a simulation
by Arlt and Brandenburg (2001) designed to capture the effect.
Clearly, thinking of ways out of
the constraints implied by helicity conservation
will be crucial to understand galactic magnetism.

\section*{Acknowledgements}
I thank Axel Brandenburg and S. Sridhar for insightful comments.

\label{lastpage}
\end{document}